\pdfoutput=1
\documentclass[traditabstract]{aa}
\usepackage{natbib}
\usepackage{txfonts}
\usepackage{graphicx}
\titlerunning{Period change \& evolution of $\beta$ Cephei stars}
\authorrunning{H. R. Neilson \& R.~Ignace}

\begin{document}

\title{Period change and stellar evolution of $\beta$ Cephei stars}

\author{Hilding R. Neilson\inst{1} \and Richard Ignace\inst{2}}
\institute{
     Department of Astronomy \& Astrophysics, University of Toronto, 50 St.~George Street, Toronto, ON, M5S~3H4, Canada \\
     \email{neilson@astro.utoronto.ca}
             \and
   Department of Physics \& Astronomy, East Tennessee State University, Box 70652, Johnson City, TN 37614 USA
}

\date{}

\abstract{
The $\beta$ Cephei stars represent an important class of massive star pulsators probing the evolution of B-type stars and the transition from main sequence to hydrogen-shell burning evolution.
By understanding $\beta$~Cep stars, we gain insights into the detailed physics of massive star evolution such as rotational mixing,  convective core overshooting, magnetic fields and stellar winds, all of which play important roles.  Similarly, modeling their pulsation provides additional information into their interior structures.  Furthermore, measurements of the rate of change of pulsation period offer a direct measure of $\beta$~Cephei stellar evolution. 
In this work, we compute state-of-the-art stellar evolution models assuming different amounts of initial rotation and convective core overshoot and measure theoretical rates of period change for which we compare to rates previously measured for a  sample of $\beta$ Cephei stars.
The results of this comparison are mixed.  For three stars, the rates are too small to infer any information from stellar evolution models, whereas for three other stars the rates are too large.  We infer stellar parameters, such as mass and age, for two $\beta$~Cephei stars: $\xi^1$~CMa and $\delta$~Cet, that agree well with independent measurements.  
 We explore ideas for why models may not predict the larger rates of period change.  In particular, period drifts in $\beta$~Cep stars can artificially lead to overestimated rates of secular period change.    
}

\keywords{Stars: evolution / Stars: fundamental parameters / Stars: mass loss / Stars: rotation}

\maketitle

\section{Introduction}
Massive B-type stars are powerful engines which drive cosmic evolution from star formation to galaxy evolution to chemical enrichment and the seeds of life in the Universe.  These stars impact star formation through their winds contributing momentum and turbulence into the interstellar medium  and chemically enrich the ISM through their supernovae. Exploring the physics of these massive stars contributes to the understanding of cosmic evolution as well as stellar physics in general. 
Massive stars are also important probes of the physics of stellar evolution because these B-type stars display many different kinds of phenomena related to the physics or stellar rotation, stellar winds, pulsation, binarity, magnetic fields, and possibly stellar mergers \citep{Langer2012}.  For instance, the emission line stars display rapid, nearly critical, rotation, while the $\beta$~Cephei stars undergo radial and non-radial pulsation  \citep{Townsend2004, Moskalik1992}.  In this work, we will explore the $\beta$~Cephei variable stars because their pulsation allows us to study their internal structure and potentially their evolution.


The $\beta$~Cephei variable stars are particularly interesting because they display a diverse range of phenomena. Some $\beta$~Cep stars appear to have strong magnetic fields \citep{Silvester2009, Hubrig2011, Neiner2012}  while others have weak or no magnetic fields \citep{Silvester2009, Fossati2015}, some rapidly rotate \citep{Aerts1994, Handler2012}, while others rotate slowly \citep{Shultz2015}.  Most, if not all, $\beta$~Cep stars tend to have weak stellar winds \citep{Martins2005, Huenemoerder2012} and are also X-ray sources.  Recently, \cite{Oskinova2014} discovered X-ray variability in one $\beta$~Cephei star $\xi^1$~CMa over a pulsation cycle.  Understanding the physics in these stars impacts our understanding of massive star physics in general along with the evolution from the main sequence to the red supergiant stage of evolution and beyond.


While probing the physics of massive stars, the $\beta$~Cephei stars also span the boundary between the main sequence evolution and the transition to the blue supergiant stage of evolution \citep{Demarque1965, Stothers1965}.  As such, observations and models of these stars help us understand the transition and precisely where on the Hertzsprung-Russell diagram it occurs thus, again, providing opportunities to refine the input physics in stellar evolution models.

In particular, there are two aspects of stellar physics that can be tested using models of $\beta$~Cephei variable stars and observations: convective core overshooting and rotational mixing \citep{Miglio2009, Lovekin2010}.  Convective core overshooting mixes material into the stellar core from layers some fraction of  a pressure scale height above.  In stellar evolution models, mixing length theory treats convection as being fixed within the convective region such that the velocity and acceleration of a convective eddy goes to zero at the boundaries.  Convective core overshooting is an ad hoc prescription in stellar evolution codes designed to account for the fact that convective eddies will not have zero velocity at the boundary between convective and radiative regions, hence penetrating above the convective core and mixing additional hydrogen into the stellar core.  This overshooting acts to extend the main sequence life time of a massive star by create a more massive core at the end of main sequence evolution.  Rotational mixing will have similar effects on massive star evolution by meridional mixing \citep[e.g.][]{Frischknecht2010, Brott2011b}.

Thanks to asteroseismic observations of $\beta$~Cephei variables, there have been numerous measurements of the amount of convective core overshoot, parameterized as a fraction of a pressure scale height, $\alpha_{co}H_P$, as the total amount of material mixed over a main sequence lifetime. The typical amount of convective core overshoot required to match observations is about $\alpha_{co} = 0.2$ and is consistent with measurements for the $\beta$~Cep stars $\beta$~CMa \citep{Mazumdar2006} and $\delta$~Ceti \citep{Aerts2006} as well as for classical Cepheids \citep{Cassisi2011, Neilson2011, Neilson2012,
 Prada2012} and other stars.  However, measurements for other $\beta$~Cep stars require different amounts of overshoot such as $\theta$~Oph  $ \alpha_{co} = 0.4$ \citep{Briquet2007} $0.28H_P$ \citep{Lovekin2010}, HD129929 $\alpha_{co} =0.1$ \citep{Aerts2003}, and $\nu$~Eridani $\alpha_{co} =0.28$ \citep{Suarez2009}.  These measurements are obtained by comparing pulsation and evolutionary models of $\beta$~Cep stars.

The degeneracy between rotational mixing and convective core overshooting is difficult to break.  Rotational mixing can be constrained from measurements of the [N/C]  enhancement \citep{Hunter2008, Brott2011b}, but this enhancement has been detected in some slowly-rotating stars \citep{Morel2006}.  There is no obvious way to constrain the importance of one process relative to another.  The goal of this work is to compare new stellar evolution models with rotation to rates of period change measured for Galactic $\beta$~Cephei stars and test whether period change is a probe of rotation and overshoot.

\cite{Eggleton1973} first compared stellar evolution models with period change measurements and determined that the rate of period change is sensitive to the stage of evolution, be it main sequence or  blue supergiant.  However, since that work, stellar evolution models have matured, incorporating new opacities plus improved prescriptions for physics of stellar mass loss, rotation and magnetic fields \citep{Yoon2005}.  For example, these changes have led to the realization that $\beta$~Cephei pulsation is driven by the iron opacity bump \citep{Moskalik1992, Cox1992, Dziembowski1993} as opposed to helium ionization in classical Cepheid and RR~Lyrae stars. With the advances in stellar evolution models, it is an opportune time to revisit the comparison with period change measurements.




In the next section, we discuss the stellar evolution models used in this work, and compute theoretical rates of pulsation period change in Sect.~3.  In Sect.~4, we compare the rates of period change and test how period change relates to rotation rates.  In Sect.~5, we discuss and summarize the results.

\section{Models}
We compute stellar evolution models using the \cite{Yoon2005} code for masses ranging from $M = 7$ to $20~M_\odot$ in steps of one solar mass with various assumptions for internal mixing.  One grid of models is computed assuming moderate convective core overshooting at $\alpha_{co} =0.2$ \citep[as described by for example][]{Huang1983, Neilson2011}   and with initial rotation rates $v_{\rm{rot}} = 100, 200$ and $400~$km~s$^{-1}$.  We include rotational mixing in these models based on the prescription of \cite{Heger2000}, which is described in more detail by \cite{Brott2011a, Brott2011b}.   A second grid of models is computed without rotation but assuming convective overshooting parameters $\alpha_{co} = 0, 0.2,$ and $0.4$.  

\begin{figure*}[t]
\begin{center}
\includegraphics[width=0.5\textwidth]{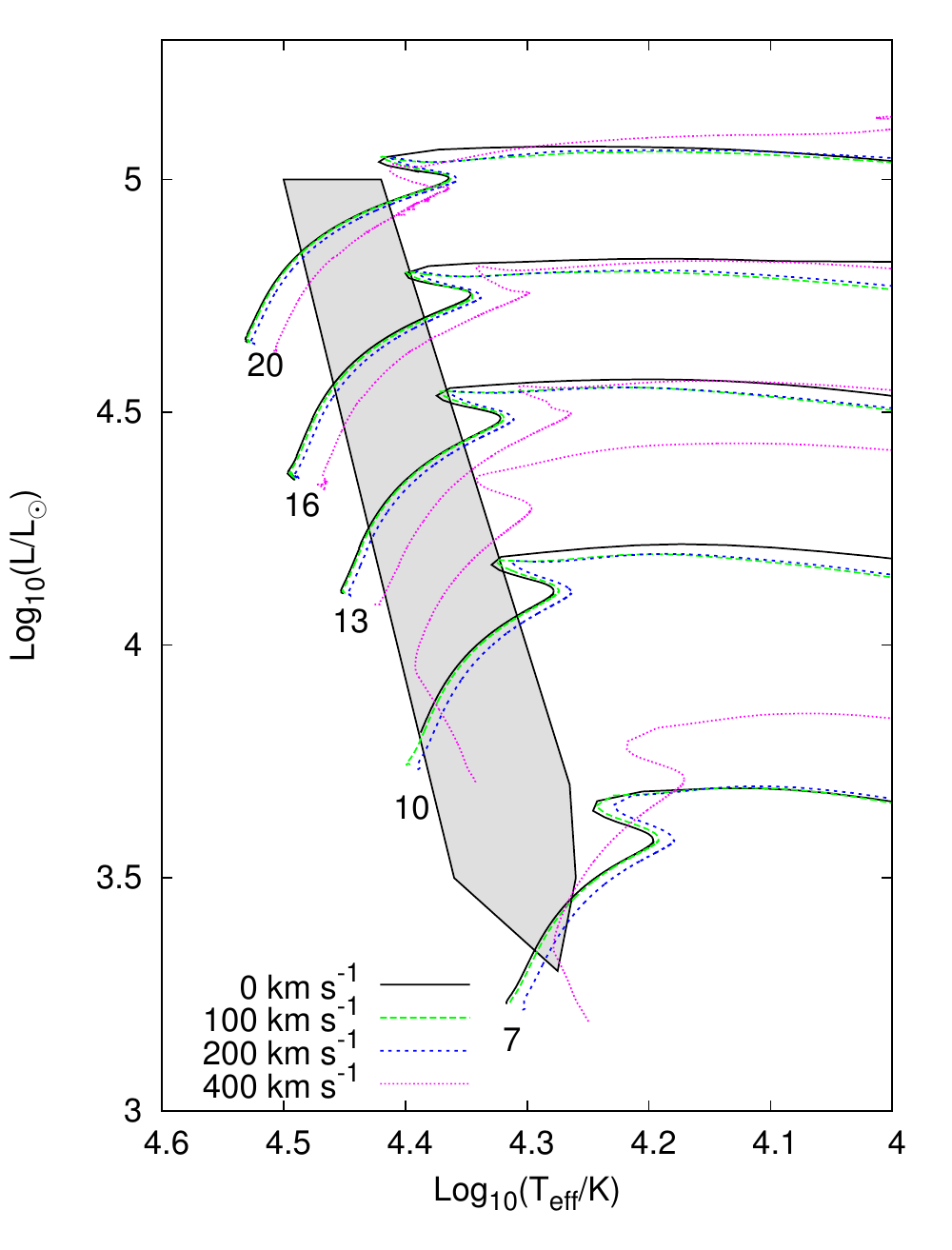}\includegraphics[width=0.5\textwidth]{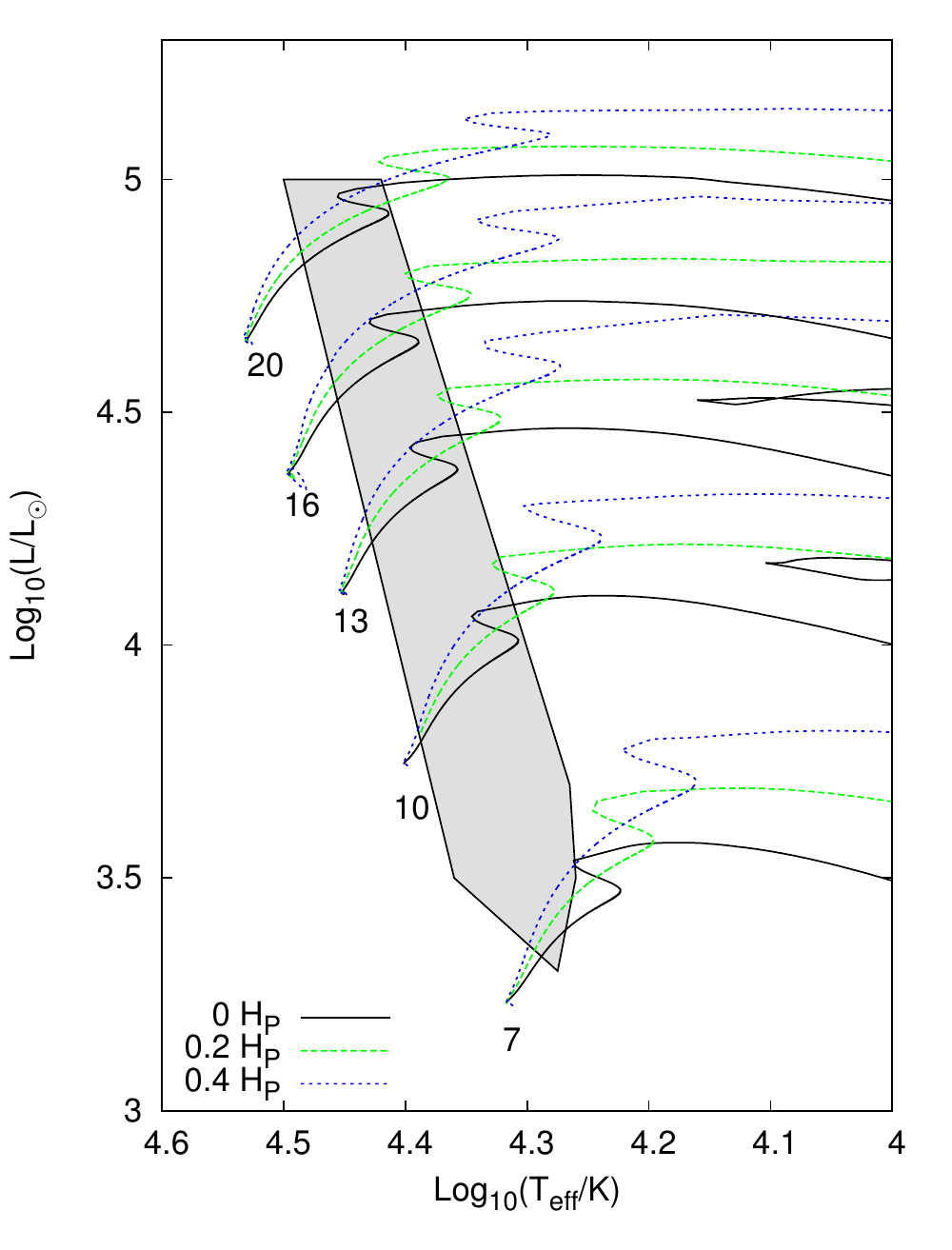}
\end{center}
\caption{(Left) Stellar evolution tracks assuming moderate convective core overshoot, $\alpha_c = 0.2$ and initial rotation rates of $0, 100, 200,$ and $400$~km~s$^{-1}$ denoted by black solid, green dashed, blue dot-dashed and magenta dotted lines, respectively.  (Right) Stellar evolution tracks assuming zero rotation and convective core overshoot, $\alpha_c = 0, 0.2$, and $0.4$ as denoted by black solid, green dashed and blue dot-dashed lines, respectively.}\label{HRD}
\end{figure*}

The stellar evolution models are computed assuming radiatively-driven winds as prescribed by \cite{Vink2000} during main sequence and blue supergiant evolution and with a \cite{Grevesse1998} solar metallicity.  We compute the evolution models from the zero-age main sequence past the boundaries of the $\beta$~Cephei instability strip as measured by \cite{Pamyatnykh2007}.  Sample stellar evolution tracks are plotted in Fig.~\ref{HRD} for different rotation rates and convective core overshooting parameters.

\section{Period Change}
The purpose of this work is to compare measured rates of period change from observations with those predicted from stellar evolution models representing $\beta$~Cephei stars.  To do so, we use an analytic model of period change based on the period-mean density relation \citep{Eddington1918},  $P\sqrt{\rho} = Q$, similar to that of \cite{Eggleton1973}. We do not use the derivations from \cite{Neilson2012a, Neilson2012b} because there is no evidence that the pulsation constant, $Q$ varies as a function of period for $\beta$~Cephei variables.  Therefore, we write the relative rate of period change as
\begin{equation}
\frac{\dot{P}}{P} = -\frac{1}{2} \frac{\dot{M}}{M} + \frac{3}{2}\frac{\dot{R}}{R},
\end{equation}
where $\dot{M}$ the rate of change of stellar mass.  This relation measures the evolution of the star as a function of time.  

\begin{figure*}[t]
\begin{center}
\includegraphics[width=0.5\textwidth]{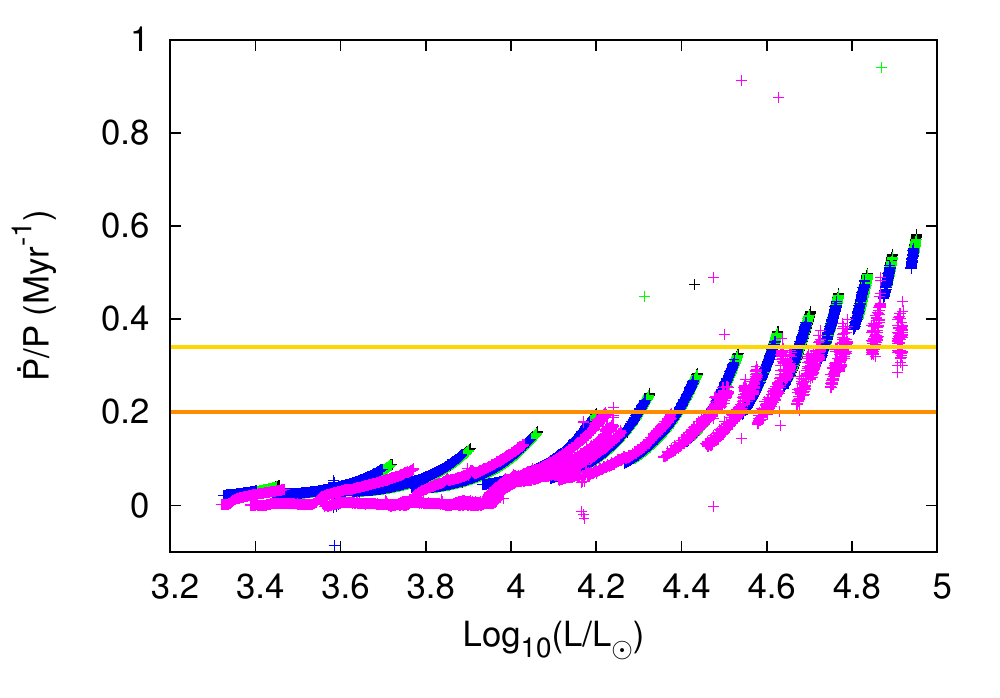}\includegraphics[width=0.5\textwidth]{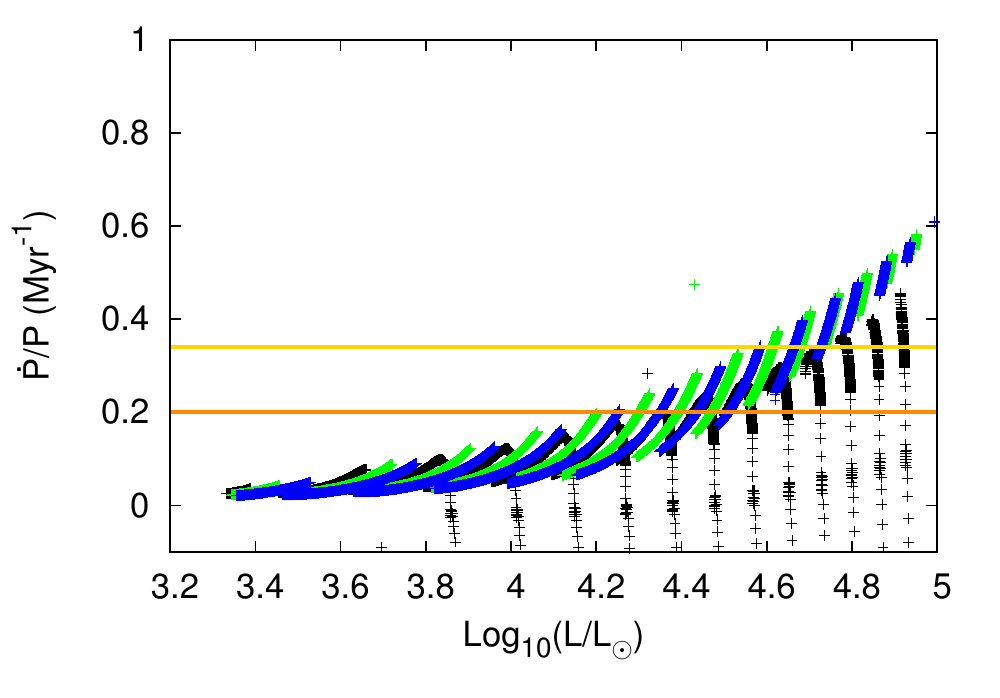}
\end{center}
\caption{(Left) Rates of period change computed from stellar evolution tracks assuming moderate convective core overshoot, $\alpha_c = 0.2$ and initial rotation rates of $0, 100, 200,$ and $400$~km~s$^{-1}$.  (Right) Rates of period change computed from stellar evolution tracks assuming zero rotation and convective core overshoot, $\alpha_c = 0, 0.2$, and $0.4$. The different colors are same as for Fig.~\ref{HRD} while the horizontal orange line denotes the measured rates of period change for $\xi^1$~CMa and $\delta$~Cet and the gold line denotes the second measured value for $\delta$~Cet.}\label{pdot}
\end{figure*}

We computed relative rates of period change from the stellar evolution models using this relation. Rates of period change are shown in Fig.~\ref{pdot} for two sets of models. The first set of models are computed assuming zero initial rotation, but with different amounts  convective core overshooting, while the second set of models assume  moderate convective core overshooting only but varying the initial rotation rates.    This result suggests that the combination of measured rates of period change and a fundamental stellar parameter such as stellar radius, effective temperature or luminosity cannot provide a meaningful measure of physical processes such as convective core overshooting and/or rotational mixing \citep[e.g.][]{Miglio2009}.  These mixing processes act to change the mean structure of the star, but only in a way that is degenerate with models with different stellar masses but no additional mixing.

As part of this analysis, we use rates of period change presented previously by \cite{Jerzykiewicz1999b} and \cite{Jerzykiewicz1999} for eight $\beta$~Cephei stars, see Table~\ref{t1}.  It should be noted that these rates differ from those presented by \cite{Eggleton1973}, based on different analyses.  Measuring period changes for $\beta$ Cephei stars is a challenge as for many of these stars there are multiple oscillation frequencies and the periods might confuse the measurement.  A second issue has been addressed by \cite{Pigulski1992, Pigulski1993,Pigulski1994}, for which binary companions affect measurements of period change due to the light time effect.  While this is not a complete sample of period change measurements for $\beta$~Cephei stars, it does represent a well-measured sample.

\begin{table}[t]
\caption{Rates of period change for a sample of $\beta$~Cep stars from \cite{Jerzykiewicz1999}.}\label{t1}
\begin{center}
\begin{tabular}{lcc}
\hline
Star & Period (d) & $\dot{P}/{P}~($Myr$^{-1})$ \\
\hline
$\gamma$ Peg & 0.15175 & $+0.046 \pm 0.015$ or $0.0 \pm 0.11$ \\
$\delta$ Cet& 0.16114&$+0.20 \pm 0.01$ or $0.34\pm0.06$ \\
$\xi^1$ CMa& 0.20958& $+0.20 \pm 0.03$ \\
$\alpha$ Lup& 0.25985&$+0.84\pm 0.22$\\
$\sigma$ Sco& 0.24684&$+1.55\pm 0.14$\\
V2052 Oph& 0.13989&$-0.10 \pm 0.22$\\
BW Vul& 0.20104&$+1.35\pm 0.02$\\
$\beta$ Cep& 0.19049&$+0.06 \pm0.06$\\
\hline
\end{tabular}
\end{center}
\end{table}

\section{Results}
We compare theoretical and measured rates of period changes for the  sample of $\beta$~Cep stars in Table~\ref{t1}. From these measurements, we can compute stellar fundamental parameters for models with rates of period change consistent with each observation.  Based on the rates presented in Table~\ref{t1} and the  models plotting in Fig.~\ref{pdot}, we break the comparisons into three categories: 1) period change rates fit by almost all models, 2) period change rates fit by specific models and 3) period change rates fit by \emph{no} models.  The first category simply suggests that the measured rate of period change offers no constraints on the models, while the third category suggests that the period change is completely inconsistent with models. These differences appear to be a result of insufficient precision for some observations and that observations also span a greater range of period changes rates than predicted from the stellar evolution models.

There are three $\beta$~Cep stars for which the measured rates of period change offer no constraints: $\gamma$~Peg, V2052~Oph, and the prototype $\beta$~Cep.  These three stars have both the smallest values for the rate of period change but also with the greatest error bars.  Similar, the stars that fall into the third category include those with the largest rates of period change: $\alpha$~Lup, $\sigma$~Sco, and BW~Vul.  While the  \cite{Jerzykiewicz1999} sample consists of only eight stars (nine measured rates of period change), there are significant hints that  there is something amiss in stellar evolution models that was not found by \cite{Eggleton1973}, even though their models did predict large rates of period change.  This is because their models  predict that stars evolve as blue supergiant stars within the $\beta$~Cephei instability strip.  All of the $\beta$~Cephei stars in the sample were classified as either luminosity class III or IV.  The three $\beta$~Cep stars with large rates of period change are luminosity class III but so is the prototype $\beta$~Cep that has a measured rate of period change consistent with zero. There is no obvious connection between rates of period change and evolutionary state of the stars.

Is there anything special about the three $\beta$~Cephei stars: $\alpha$~Lup, $\sigma$~Sco, and BW~Vul? These three stars appear to pulsate consistently in the fundamental mode, but there is nothing apparently unique about these stars relative to the others in the sample.  The mystery deepens when one notes that the other $\beta$~Cephei stars  in our sample tend to have magnetic fields, which are not modelled in this work, and many are slow rotators.  The stars that are consistent with none of the stellar evolution models are the ones that are expected to be most easily fit. 

The three stars with rates of period change that are consistent with predictions from all stellar evolution models, $\gamma$~Peg, V2052~Oph, and $\beta$~Cep, also provide no clear  answers.  There have been claims that  $\gamma$~Peg has a weak magnetic field \citep{Butkovskaya2007}, but \cite{Neiner2014} reported a  null measurement.  The star also has a measured rotation rate of about 3~km~s$^{-1}$ \citep{Telting2006, Handler2009}.  These observations suggest the stellar evolution models computed in this work should be sufficient for analyzing the stellar properties.  On the other hand, V2052~Oph is a magnetic He-strong star  \citep{Neiner2012}.  \cite{Briquet2012} measured rotation rates of about 70 -- 75~km~s$^{-1}$ and smaller amounts of convective core overshooting than for other $\beta$~Cephei variable stars.  The prototype $\beta$~Cep is similarly a magnetic He-strong star with a varying magnetic field of about 100~$G$ \citep{Henrichs2013}. The three $\beta$~Cep stars that cannot be constrained, i.e., consistent with {\it all} of the computed models, appear to have little in common.

Of the two stars for which we find rates of period change that can measure stellar properties: $\xi^1$~CMa and $\delta$~Cet,  the former has a measured magnetic field strength of about 100~$G$ \citep{Hubrig2006, Silvester2009} and very slow rotation. This behavior is consistent with spin down due to the magnetic field interactions \citep{Shultz2015}.  There have been no measurements  of magnetic fields in $\delta$~Cet, but it does have a nitrogen excess and slow rotation rate consistent with known magnetic $\beta$~Cep stars \citep{Morel2006,  Levenhagen2013}. 

In the next subsections, we explore in more detail the predicted stellar properties for $\xi^1$~CMa and $\delta$~Cet in combination with measurements of rotation rates and measurements of fundamental parameters from seismic and other analyses.
\begin{table}[t]
\caption{Measured fundamental properties of two $\beta$~Cephei stars from the literature.}\label{t2}
\begin{center}
\begin{tabular}{lcc}
Property & $\xi^1$~CMa$^{\rm{a}}$ & $\delta$~Cet$^{\rm{b}}$ \\
\hline
Mass ($M_\odot$)& 13 -- 15 & 7.9 \\
$T_{eff}$ (K) & $25900 \pm 100$ & $21675\pm 400$ \\
Radius ($R_\odot$) & $8.7\pm 0.7$ & --- \\
$\log L/L_\odot$ & $4.46\pm 0.07$ & 3.6 \\
\hline
\end{tabular}
\note{(a) \cite{Shultz2015}, (b) \cite{Levenhagen2013}}
\end{center}
\end{table}

\subsection{Properties of $\xi^1$~CMa}
\cite{Shultz2015} presented a Baade-Wesselink analysis of $\xi^1$~CMa using Hipparcos photometry and radial velocity measurements \citep{Saesen2006}.  They measure a mean effective temperature of $25.9\pm 0.1$~kK, a radius $R = 8.7\pm 0.7~R_\odot$ and a luminosity $\log L/L_\odot = 4.46\pm 0.07$.  The authors also measure the age of $\xi^1$~CMa to be about 12.6~Myr from isochrones \citep{Ekstrom2012}. Furthermore, they note that the star is rotating slowly with a period of about 60 years.

Using the measured rate of period change and our stellar evolution models assuming zero rotation only we measure an age between 8 -- 18 Myr, consistent with results from the \cite{Ekstrom2012} stellar isochrones.  However, when we include the effective temperature as an additional constraint of the stellar parameters, we measure a luminosity between $\log L/L_\odot = 4.40$ -- $4.60$, but with a younger age of 8 -- 12 Myr.  Furthermore, we measure the mass of $\xi^1~$CMa to be $M = 13$ -- $15~M_\odot$ with a small probability of it being more massive if there is no convective core overshooting in the star.   

This comparison is striking because of how consistent the measured rate of period change is with predictions from our models while for other cases  predicted rates are too small relative to observations.  Another issue is that $\xi^1$~CMa has a strong magnetic field, which is not included in our models.  If the magnetic field is strong enough then one might expect it to affect the pulsation properties of the star, but these findings suggest the magnetic field plays no role in the period change.  It is not clear why the models provide such consistent predictions of stellar parameters for this star but not for others \citep{Mathis2015}.

\subsection{Properties of $\delta$~Cet}
The second case where predicted rates of period change appear to constrain stellar fundamental parameters is for $\delta$~Cet.  \cite{Levenhagen2013} fit spectroscopic observations with non-LTE models and measured $T_{\rm{eff}} = 21675\pm 400$~K and $\log g = 4.03\pm 0.08$.  Other reports measure similar values for the effective temperature but gravities as small as $\log g = 3.7$ \citep{Niemczura2005, Morel2006, Hubrig2009}.  \cite{Levenhagen2013} also measured a projected rotation velocity of about $v\sin i = 13$~km~s$^{-1}$, suggesting that the star is slowly rotating or is being viewed nearly pole-on.  Comparing their results to stellar evolution models, they also infer a mass $= 7.9~M_\odot$ and a luminosity of $\log L/L_\odot = 3.6$.

When we compare our stellar evolution models to both measurements of period change for $\delta$~Cet, we find weakly constrained values of stellar parameters.  The predicted age, from 7 -- 18~Myr, is consistent with previous results  \citep{Levenhagen2013}. The effective temperature is predicted to range from $\log T_{\rm{eff}} = 4.32$ -- $4.42$, and the luminosity is $\log L/L_\odot = 4.2$ -- $5$.  The effective temperature is consistent with observations but the luminosity is significantly greater. There is no improvement if we consider only stellar evolution models with no rotation as opposed to the case of $\xi^1$~CMa. 

If we confine our models to only those with effective temperatures consistent with the \cite{Levenhagen2013} measurements, we better constrain the other stellar parameters.  In this case, we measure the mass of $\delta$~Cet to be $M = 10$ -- $12~M_\odot$ and its luminosity is $\log L/L_\odot  = 4.18$ -- $4.25$.  The luminosity is still significantly greater than that measured by \cite{Levenhagen2013} using \cite{Schaller1992} stellar evolution tracks.  The differences could be due to different prescriptions of interior mixing, opacities and input physics, making it unlikely that our results indicate some unknown underlying issue with the stellar evolution models.

While a magnetic field has not been directly measured for $\delta$~Cet, the star shows nitrogen enhancement which is suggestive of one \citep{Morel2006, Levenhagen2013}.  When coupled with the apparent slow rotation of $\delta$~Cet, it appears likely that $\delta$~Ceti is similar to $\xi^{1}$~CMa. This result further raises questions as to why stellar evolution models appear to fit these two stars, but not the non-magnetic  $\beta$~Cephei variables in the sample.

\section{Discussion}
The goal of our work was to revisit the results of \cite{Eggleton1973} comparing stellar evolution models and predicted rates of period change with observations.  \cite{Eggleton1973} found that stellar evolution models fit observed rates of period change for a sample of $\beta$~Cephei stars well and could explain the large range of observed rates in terms of both hydrogen-core and hydrogen-shell burning stars.  While this evolutionary scenario explains the diversity of period change measurements in $\beta$~Cephei variable stars, stellar evolution models have advanced significantly in the past forty years. 

Since the work of \cite{Eggleton1973}, stellar opacities have changed and  new physical processes have been added to stellar evolution models.  We computed grids of models at solar composition, for masses from $M = 7$ -- $20~M_\odot$.  The models included various amounts of convective core overshooting along with different amounts of initial rotation. For every model, we compute relative rates of period change when the stars evolve along the $\beta$~Cephei instability strip. Both rotation and overshooting extend the main sequence lifetimes of stars, hence act to decrease the relative rates of period change. As such, comparing predicted rates of period change to observations do not allow us to measure the rotation history of a star or the amounts of overshooting because it is not possible to disentangle the stellar mass from the amount of mixing in the stellar evolution models.



We compared our predicted rates to period change measurements for a sample of eight stars \citep{Jerzykiewicz1999}, but found mixed results.  For three of the stars: $\gamma$~Peg, V2052~Oph and the prototype $\beta$~Cep, our models could not be applied to constrain any fundamental stellar parameters.  The rates of period change for these stars are all consistent  with being constant and there are points in the evolution of every star  along the $\beta$~Cephei instability strip where the rate of period change is  very small. For three other stars: BW~Vul, $\sigma$~Sco, and $\alpha$~Lup, the predicted rates of period change for all stellar evolution models are too small relative to the measured rates.
 For two stars $\xi^1$~CMa and $\delta$~Cet, we predicted fundamental stellar parameters by comparing predicted and observed rates of period change.  The predicted stellar parameters are consistent with previous independent measurements. We also measure the masses for $\delta$~Cet and $xi^1$~CMa to be $M = 11 \pm 1~M_\odot$ and $M  = 14\pm 1~M_\odot$, respectively. 

The fits to the measured rates of period change of $\xi^1$~CMa and $\delta$~Cet are both reassuring and surprising.  That we find agreement with our models suggests that the analysis is consistent and the models are good representations of real stars. However, there is evidence for strong magnetic fields in the two stars, but these are not taken into account for the stellar evolution models.  That the models tend to agree suggests that the presence of a strong magnetic field might not play a significant role in the pulsation of these stars relative to secular stellar evolution.

Those stars with measured rates of period change that are greater than any of those predicted by stellar evolution models is disconcerting.   No amount of rotation or convective core overshooting can resolve that difference nor do these three stars have any common property that could explain the difference.  For instance, only V2052~Oph shows evidence for a strong magnetic field.  There are a number of reasons as to why theoretical rates of period change might be different from those measured. If these three stars were undergoing mass loss that is about one or two orders of magnitude greater than predicted then the models would have rates of period change similar to that observed. However, there is no reason why  mass loss would be enhanced by such an amount for these stars but not the other five, including  $\xi^1$~CMa and $\delta$~Cet.  

We computed rates of period change by differentiating the period-mean density relation.  If $\beta$~Cephei variable star pulsation followed a different relation then the predicted rates of period change could be significantly different, but again, if we use a different relation then the agreement found for $\xi^1$~CMa and $\delta$~Cet would be lost.  

A third option is that the measured rates of period change for these stars do not represent only secular period change, but include  the effects of possible period drift  where the pulsation of the star changes abruptly over short timescales due to some non-evolutionary cause.  \cite{Odell2012} analyzed timing measurements for BW~Vul and argued that the pulsation period is mostly constant with respect to time, but underwent random fluctuations during various times in the past.  These changes are not secular, i.e., not a function of evolution, but this cause is unclear.  A similar period drift has been suspected for the classical Cepheid Polaris \citep{Turner2005, Neilson2012a}.  If this period drift is occurring in $\beta$ Cephei variable stars then it could explain why there are such large rates of period change for only three stars.  It could also explain the two $\beta$~Cephei variable stars for which the rates of period change are well-fit, the stars have not undergone period drifts over the time scales of the observations. 

Our results suggest the evolution of $\beta$~Cephei variable stars and their pulsation properties are more complex than expected, raising questions about their rates of period change.  However, our results are based on measurements for only eight stars, necessitating period change measurements for more $\beta$~Cephei variable stars along with continued observations of the eight stars discussed in  \cite{Jerzykiewicz1999}.    Similarly, our models assume that the period-mean density relation is constant and is a valid measure of secular period change that should be verified by direct modelling of pulsation and evolution.

\bibliographystyle{aa} 

\bibliography{beta_cep} 

\begin{thebibliography}{64}
\expandafter\ifx\csname natexlab\endcsname\relax\def\natexlab#1{#1}\fi

\bibitem[{{Aerts} {et~al.}(2006){Aerts}, {Marchenko}, {Matthews}, {Kuschnig},
  {Guenther}, {Moffat}, {Rucinski}, {Sasselov}, {Walker}, \&
  {Weiss}}]{Aerts2006}
{Aerts}, C., {Marchenko}, S.~V., {Matthews}, J.~M., {et~al.} 2006, \apj, 642,
  470

\bibitem[{{Aerts} {et~al.}(2003){Aerts}, {Thoul}, {Daszy{\'n}ska}, {Scuflaire},
  {Waelkens}, {Dupret}, {Niemczura}, \& {Noels}}]{Aerts2003}
{Aerts}, C., {Thoul}, A., {Daszy{\'n}ska}, J., {et~al.} 2003, Science, 300,
  1926

\bibitem[{{Aerts} {et~al.}(1994){Aerts}, {Waelkens}, \& {de Pauw}}]{Aerts1994}
{Aerts}, C., {Waelkens}, C., \& {de Pauw}, M. 1994, \aap, 286, 136

\bibitem[{{Briquet} {et~al.}(2007){Briquet}, {Morel}, {Thoul}, {Scuflaire},
  {Miglio}, {Montalb{\'a}n}, {Dupret}, \& {Aerts}}]{Briquet2007}
{Briquet}, M., {Morel}, T., {Thoul}, A., {et~al.} 2007, \mnras, 381, 1482

\bibitem[{{Briquet} {et~al.}(2012){Briquet}, {Neiner}, {Aerts}, {Morel},
  {Mathis}, {Reese}, {Lehmann}, {Costero}, {Echevarria}, {Handler}, {Kambe},
  {Hirata}, {Masuda}, {Wright}, {Yang}, {Pintado}, {Mkrtichian}, {Lee}, {Han},
  {Bruch}, {De Cat}, {Uytterhoeven}, {Lefever}, {Vanautgaerden}, {de Batz},
  {Fr{\'e}mat}, {Henrichs}, {Geers}, {Martayan}, {Hubert}, {Thizy}, \&
  {Tijani}}]{Briquet2012}
{Briquet}, M., {Neiner}, C., {Aerts}, C., {et~al.} 2012, \mnras, 427, 483

\bibitem[{{Brott} {et~al.}(2011{\natexlab{a}}){Brott}, {de Mink}, {Cantiello},
  {Langer}, {de Koter}, {Evans}, {Hunter}, {Trundle}, \& {Vink}}]{Brott2011a}
{Brott}, I., {de Mink}, S.~E., {Cantiello}, M., {et~al.} 2011{\natexlab{a}},
  \aap, 530, A115

\bibitem[{{Brott} {et~al.}(2011{\natexlab{b}}){Brott}, {Evans}, {Hunter}, {de
  Koter}, {Langer}, {Dufton}, {Cantiello}, {Trundle}, {Lennon}, {de Mink},
  {Yoon}, \& {Anders}}]{Brott2011b}
{Brott}, I., {Evans}, C.~J., {Hunter}, I., {et~al.} 2011{\natexlab{b}}, \aap,
  530, A116

\bibitem[{{Butkovskaya} \& {Plachinda}(2007)}]{Butkovskaya2007}
{Butkovskaya}, V.~V. \& {Plachinda}, S.~I. 2007, \aap, 469, 1069

\bibitem[{{Cassisi} \& {Salaris}(2011)}]{Cassisi2011}
{Cassisi}, S. \& {Salaris}, M. 2011, \apjl, 728, L43

\bibitem[{{Cox} {et~al.}(1992){Cox}, {Morgan}, {Rogers}, \&
  {Iglesias}}]{Cox1992}
{Cox}, A.~N., {Morgan}, S.~M., {Rogers}, F.~J., \& {Iglesias}, C.~A. 1992,
  \apj, 393, 272

\bibitem[{{Demarque} \& {Percy}(1965)}]{Demarque1965}
{Demarque}, P. \& {Percy}, J.~R. 1965, \aj, 70, 136

\bibitem[{{Dziembowski} {et~al.}(1993){Dziembowski}, {Moskalik}, \&
  {Pamyatnykh}}]{Dziembowski1993}
{Dziembowski}, W.~A., {Moskalik}, P., \& {Pamyatnykh}, A.~A. 1993, \mnras, 265,
  588

\bibitem[{{Eddington}(1918)}]{Eddington1918}
{Eddington}, A.~S. 1918, \mnras, 79, 2

\bibitem[{{Eggleton} \& {Percy}(1973)}]{Eggleton1973}
{Eggleton}, P.~P. \& {Percy}, J.~R. 1973, \mnras, 161, 421

\bibitem[{{Ekstr{\"o}m} {et~al.}(2012){Ekstr{\"o}m}, {Georgy}, {Eggenberger},
  {Meynet}, {Mowlavi}, {Wyttenbach}, {Granada}, {Decressin}, {Hirschi},
  {Frischknecht}, {Charbonnel}, \& {Maeder}}]{Ekstrom2012}
{Ekstr{\"o}m}, S., {Georgy}, C., {Eggenberger}, P., {et~al.} 2012, \aap, 537,
  A146

\bibitem[{{Fossati} {et~al.}(2015){Fossati}, {Castro}, {Morel}, {Langer},
  {Briquet}, {Carroll}, {Hubrig}, {Nieva}, {Oskinova}, {Przybilla},
  {Schneider}, {Sch{\"o}ller}, {Sim{\'o}n-D{\'{\i}}az}, {Ilyin}, {de Koter},
  {Reisenegger}, \& {Sana}}]{Fossati2015}
{Fossati}, L., {Castro}, N., {Morel}, T., {et~al.} 2015, \aap, 574, A20

\bibitem[{{Frischknecht} {et~al.}(2010){Frischknecht}, {Hirschi}, {Meynet},
  {Ekstr{\"o}m}, {Georgy}, {Rauscher}, {Winteler}, \&
  {Thielemann}}]{Frischknecht2010}
{Frischknecht}, U., {Hirschi}, R., {Meynet}, G., {et~al.} 2010, \aap, 522, A39

\bibitem[{{Grevesse} \& {Sauval}(1998)}]{Grevesse1998}
{Grevesse}, N. \& {Sauval}, A.~J. 1998, \ssr, 85, 161

\bibitem[{{Handler}(2009)}]{Handler2009}
{Handler}, G. 2009, \mnras, 398, 1339

\bibitem[{{Handler} {et~al.}(2012){Handler}, {Shobbrook}, {Uytterhoeven},
  {Briquet}, {Neiner}, {Tshenye}, {Ngwato}, {van Winckel}, {Guggenberger},
  {Raskin}, {Rodr{\'{\i}}guez}, {Mazumdar}, {Barban}, {Lorenz},
  {Vandenbussche}, {{\c S}ahin}, {Medupe}, \& {Aerts}}]{Handler2012}
{Handler}, G., {Shobbrook}, R.~R., {Uytterhoeven}, K., {et~al.} 2012, \mnras,
  424, 2380

\bibitem[{{Heger} \& {Langer}(2000)}]{Heger2000}
{Heger}, A. \& {Langer}, N. 2000, \apj, 544, 1016

\bibitem[{{Henrichs} {et~al.}(2013){Henrichs}, {de Jong}, {Verdugo}, {Schnerr},
  {Neiner}, {Donati}, {Catala}, {Shorlin}, {Wade}, {Veen}, {Nichols}, {Damen},
  {Talavera}, {Hill}, {Kaper}, {Tijani}, {Geers}, {Wiersema}, {Plaggenborg}, \&
  {Rygl}}]{Henrichs2013}
{Henrichs}, H.~F., {de Jong}, J.~A., {Verdugo}, E., {et~al.} 2013, \aap, 555,
  A46

\bibitem[{{Huang} \& {Weigert}(1983)}]{Huang1983}
{Huang}, R.~Q. \& {Weigert}, A. 1983, \aap, 127, 309

\bibitem[{{Hubrig} {et~al.}(2009){Hubrig}, {Briquet}, {De Cat}, {Sch{\"o}ller},
  {Morel}, \& {Ilyin}}]{Hubrig2009}
{Hubrig}, S., {Briquet}, M., {De Cat}, P., {et~al.} 2009, Astronomische
  Nachrichten, 330, 317

\bibitem[{{Hubrig} {et~al.}(2006){Hubrig}, {Briquet}, {Sch{\"o}ller}, {De Cat},
  {Mathys}, \& {Aerts}}]{Hubrig2006}
{Hubrig}, S., {Briquet}, M., {Sch{\"o}ller}, M., {et~al.} 2006, \mnras, 369,
  L61

\bibitem[{{Hubrig} {et~al.}(2011){Hubrig}, {Ilyin}, {Sch{\"o}ller}, {Briquet},
  {Morel}, \& {De Cat}}]{Hubrig2011}
{Hubrig}, S., {Ilyin}, I., {Sch{\"o}ller}, M., {et~al.} 2011, \apjl, 726, L5

\bibitem[{{Huenemoerder} {et~al.}(2012){Huenemoerder}, {Oskinova}, {Ignace},
  {Waldron}, {Todt}, {Hamaguchi}, \& {Kitamoto}}]{Huenemoerder2012}
{Huenemoerder}, D.~P., {Oskinova}, L.~M., {Ignace}, R., {et~al.} 2012, \apjl,
  756, L34

\bibitem[{{Hunter} {et~al.}(2008){Hunter}, {Brott}, {Lennon}, {Langer},
  {Dufton}, {Trundle}, {Smartt}, {de Koter}, {Evans}, \& {Ryans}}]{Hunter2008}
{Hunter}, I., {Brott}, I., {Lennon}, D.~J., {et~al.} 2008, \apjl, 676, L29

\bibitem[{{Jerzykiewicz}(1999)}]{Jerzykiewicz1999}
{Jerzykiewicz}, M. 1999, \nar, 43, 455

\bibitem[{{Jerzykiewicz} \& {Pigulski}(1999)}]{Jerzykiewicz1999b}
{Jerzykiewicz}, M. \& {Pigulski}, A. 1999, \nar, 43, 461

\bibitem[{{Langer}(2012)}]{Langer2012}
{Langer}, N. 2012, \araa, 50, 107

\bibitem[{{Levenhagen} {et~al.}(2013){Levenhagen}, {K{\"u}nzel}, \&
  {Leister}}]{Levenhagen2013}
{Levenhagen}, R.~S., {K{\"u}nzel}, R., \& {Leister}, N.~V. 2013, \na, 18, 55

\bibitem[{{Lovekin} \& {Goupil}(2010)}]{Lovekin2010}
{Lovekin}, C.~C. \& {Goupil}, M.-J. 2010, \aap, 515, A58

\bibitem[{{Martins} {et~al.}(2005){Martins}, {Schaerer}, {Hillier},
  {Meynadier}, {Heydari-Malayeri}, \& {Walborn}}]{Martins2005}
{Martins}, F., {Schaerer}, D., {Hillier}, D.~J., {et~al.} 2005, \aap, 441, 735

\bibitem[{{Mathis} \& {Neiner}(2015)}]{Mathis2015}
{Mathis}, S. \& {Neiner}, C. 2015, in IAU Symposium, Vol. 307, IAU Symposium,
  420--425

\bibitem[{{Mazumdar} {et~al.}(2006){Mazumdar}, {Briquet}, {Desmet}, \&
  {Aerts}}]{Mazumdar2006}
{Mazumdar}, A., {Briquet}, M., {Desmet}, M., \& {Aerts}, C. 2006, \aap, 459,
  589

\bibitem[{{Miglio} {et~al.}(2009){Miglio}, {Montalb{\'a}n}, {Eggenberger}, \&
  {Noels}}]{Miglio2009}
{Miglio}, A., {Montalb{\'a}n}, J., {Eggenberger}, P., \& {Noels}, A. 2009,
  Communications in Asteroseismology, 158, 233

\bibitem[{{Morel} {et~al.}(2006){Morel}, {Butler}, {Aerts}, {Neiner}, \&
  {Briquet}}]{Morel2006}
{Morel}, T., {Butler}, K., {Aerts}, C., {Neiner}, C., \& {Briquet}, M. 2006,
  \aap, 457, 651

\bibitem[{{Moskalik} \& {Dziembowski}(1992)}]{Moskalik1992}
{Moskalik}, P. \& {Dziembowski}, W.~A. 1992, \aap, 256, L5

\bibitem[{{Neilson} {et~al.}(2011){Neilson}, {Cantiello}, \&
  {Langer}}]{Neilson2011}
{Neilson}, H.~R., {Cantiello}, M., \& {Langer}, N. 2011, \aap, 529, L9

\bibitem[{{Neilson} {et~al.}(2012{\natexlab{a}}){Neilson}, {Engle}, {Guinan},
  {Langer}, {Wasatonic}, \& {Williams}}]{Neilson2012a}
{Neilson}, H.~R., {Engle}, S.~G., {Guinan}, E., {et~al.} 2012{\natexlab{a}},
  \apjl, 745, L32

\bibitem[{{Neilson} \& {Langer}(2012)}]{Neilson2012}
{Neilson}, H.~R. \& {Langer}, N. 2012, \aap, 537, A26

\bibitem[{{Neilson} {et~al.}(2012{\natexlab{b}}){Neilson}, {Langer}, {Engle},
  {Guinan}, \& {Izzard}}]{Neilson2012b}
{Neilson}, H.~R., {Langer}, N., {Engle}, S.~G., {Guinan}, E., \& {Izzard}, R.
  2012{\natexlab{b}}, \apjl, 760, L18

\bibitem[{{Neiner} {et~al.}(2012){Neiner}, {Alecian}, {Briquet}, {Floquet},
  {Fr{\'e}mat}, {Martayan}, {Thizy}, \& {Mimes Collaboration}}]{Neiner2012}
{Neiner}, C., {Alecian}, E., {Briquet}, M., {et~al.} 2012, \aap, 537, A148

\bibitem[{{Neiner} {et~al.}(2014){Neiner}, {Monin}, {Leroy}, {Mathis}, \&
  {Bohlender}}]{Neiner2014}
{Neiner}, C., {Monin}, D., {Leroy}, B., {Mathis}, S., \& {Bohlender}, D. 2014,
  \aap, 562, A59

\bibitem[{{Niemczura} \& {Daszy{\'n}ska-Daszkiewicz}(2005)}]{Niemczura2005}
{Niemczura}, E. \& {Daszy{\'n}ska-Daszkiewicz}, J. 2005, \aap, 433, 659

\bibitem[{{Odell}(2012)}]{Odell2012}
{Odell}, A.~P. 2012, \aap, 544, A28

\bibitem[{{Oskinova} {et~al.}(2014){Oskinova}, {Naz{\'e}}, {Todt},
  {Huenemoerder}, {Ignace}, {Hubrig}, \& {Hamann}}]{Oskinova2014}
{Oskinova}, L.~M., {Naz{\'e}}, Y., {Todt}, H., {et~al.} 2014, Nature
  Communications, 5, 4024

\bibitem[{{Pamyatnykh}(2007)}]{Pamyatnykh2007}
{Pamyatnykh}, A.~A. 2007, Communications in Asteroseismology, 150, 207

\bibitem[{{Pigulski}(1992)}]{Pigulski1992}
{Pigulski}, A. 1992, \aap, 261, 203

\bibitem[{{Pigulski}(1993)}]{Pigulski1993}
{Pigulski}, A. 1993, \aap, 274, 269

\bibitem[{{Pigulski}(1994)}]{Pigulski1994}
{Pigulski}, A. 1994, \aap, 292, 183

\bibitem[{{Prada Moroni} {et~al.}(2012){Prada Moroni}, {Gennaro}, {Bono},
  {Pietrzy{\'n}ski}, {Gieren}, {Pilecki}, {Graczyk}, \& {Thompson}}]{Prada2012}
{Prada Moroni}, P.~G., {Gennaro}, M., {Bono}, G., {et~al.} 2012, \apj, 749, 108

\bibitem[{{Saesen} {et~al.}(2006){Saesen}, {Briquet}, \& {Aerts}}]{Saesen2006}
{Saesen}, S., {Briquet}, M., \& {Aerts}, C. 2006, Communications in
  Asteroseismology, 147, 109

\bibitem[{{Schaller} {et~al.}(1992){Schaller}, {Schaerer}, {Meynet}, \&
  {Maeder}}]{Schaller1992}
{Schaller}, G., {Schaerer}, D., {Meynet}, G., \& {Maeder}, A. 1992, \aaps, 96,
  269

\bibitem[{{Shultz} {et~al.}(2015){Shultz}, {Wade}, {Rivinius}, {Marcolino},
  {Henrichs}, \& {Grunhut}}]{Shultz2015}
{Shultz}, M., {Wade}, G., {Rivinius}, T., {et~al.} 2015, in IAU Symposium, Vol.
  307, IAU Symposium, 399--400

\bibitem[{{Silvester} {et~al.}(2009){Silvester}, {Neiner}, {Henrichs}, {Wade},
  {Petit}, {Alecian}, {Huat}, {Martayan}, {Power}, \& {Thizy}}]{Silvester2009}
{Silvester}, J., {Neiner}, C., {Henrichs}, H.~F., {et~al.} 2009, \mnras, 398,
  1505

\bibitem[{{Stothers}(1965)}]{Stothers1965}
{Stothers}, R. 1965, \apj, 141, 671

\bibitem[{{Su{\'a}rez} {et~al.}(2009){Su{\'a}rez}, {Moya}, {Amado},
  {Mart{\'{\i}}n-Ruiz}, {Rodr{\'{\i}}guez-L{\'o}pez}, \&
  {Garrido}}]{Suarez2009}
{Su{\'a}rez}, J.~C., {Moya}, A., {Amado}, P.~J., {et~al.} 2009, \apj, 690, 1401

\bibitem[{{Telting} {et~al.}(2006){Telting}, {Schrijvers}, {Ilyin},
  {Uytterhoeven}, {De Ridder}, {Aerts}, \& {Henrichs}}]{Telting2006}
{Telting}, J.~H., {Schrijvers}, C., {Ilyin}, I.~V., {et~al.} 2006, \aap, 452,
  945

\bibitem[{{Townsend} {et~al.}(2004){Townsend}, {Owocki}, \&
  {Howarth}}]{Townsend2004}
{Townsend}, R.~H.~D., {Owocki}, S.~P., \& {Howarth}, I.~D. 2004, \mnras, 350,
  189

\bibitem[{{Turner} {et~al.}(2005){Turner}, {Savoy}, {Derrah}, {Abdel-Sabour
  Abdel-Latif}, \& {Berdnikov}}]{Turner2005}
{Turner}, D.~G., {Savoy}, J., {Derrah}, J., {Abdel-Sabour Abdel-Latif}, M., \&
  {Berdnikov}, L.~N. 2005, \pasp, 117, 207

\bibitem[{{Vink} {et~al.}(2000){Vink}, {de Koter}, \& {Lamers}}]{Vink2000}
{Vink}, J.~S., {de Koter}, A., \& {Lamers}, H.~J.~G.~L.~M. 2000, \aap, 362, 295

\bibitem[{{Yoon} \& {Langer}(2005)}]{Yoon2005}
{Yoon}, S.-C. \& {Langer}, N. 2005, \aap, 443, 643

\end{thebibliography}

\end{document}